\documentclass[sensors,article,accept,moreauthors,pdftex]{Definitions/mdpi} 

\usepackage{makecell}

\firstpage{1} 
\pubvolume{1}
\issuenum{1}
\articlenumber{0}
\pubyear{2021}
\copyrightyear{2021}

\externaleditor{Communicated by: César Briso-Rodríguez, Ke Guan, Andrej Hrovat and Youyun Xu} 

\datereceived{15 October 2021} 
\dateaccepted{1 December 2021} 
\datepublished{} 
\hreflink{https://doi.org/} 

\Title{Unmanned Aerial Vehicle Propagation Channel over Vegetation and Lake Areas: First- and Second-Order Statistical Analysis} 

\TitleCitation{Unmanned Aerial Vehicle Propagation Channel over Vegetation and Lake Areas: First- and Second-Order Statistical Analysis} 

\Author{Deyvid L. Leite 
 $^{1,}$*\orcidA{}, Pablo Javier Alsina $^{1}$\orcidB{}, Millena M. de M. Campos $^{1}$\orcidC{}, Vicente A. de Sousa Junior 
$^{1}$\orcidD{} \mbox{and Alvaro A. M. de Medeiros} $^{2}$\orcidE{}}

\AuthorNames{Deyvid L. Leite, Pablo Javier Alsina, Millena M. de M. Campos, Vicente A. de Sousa Jr. and Alvaro A. M. de Medeiros}
\AuthorCitation{Leite, D.L.; Alsina, P.J.; Campos, M.M.d.M.; Junior, V.A.d.S.; Medeiros, A.A.M.d.}

\address{%
$^{1}$ \quad {Post-Graduate Program in Electrical and Computer Engineering, Federal University of Rio Grande do Norte (UFRN)}
, {Natal} 
, 59078-970, RN, Brazil; pablo@dca.ufrn.br (P.J.A.); millena.gppcom@gmail.com (M.M.d.M.C.); vicente.sousa@ufrn.edu.br (V.A.d.S.J.)\\
$^{2}$ \quad {Department of Electrical Engineering, Federal University of Juiz de Fora}, {Juiz de Fora}, 36036-900, MG, Brazil; alvaro@engenharia.ufjf.br
}

\corres{Correspondence: deyvid.lucas.095@ufrn.edu.br}

\abstract{The use of unmanned aerial vehicles (UAV) to provide services such as the Internet, goods delivery, and air taxis has become a reality in recent years. The use of these aircraft requires a secure communication between the control station and the UAV, which demands the characterization of the communication channel. This paper aims to present a measurement setup using an unmanned aircraft to acquire data for the characterization of the radio frequency channel in a propagation environment with particular vegetation (Caatinga) and a lake. 
This paper presents the following contributions: identification of the communication channel model that best describes the characteristics of communication; characterization of the effects of large-scale fading, such as path loss and log-normal shadowing; characterization of small-scale fading (multipath and Doppler); and estimation of the aircraft speed from the identified Doppler frequency.}


\usepackage{graphicx}
\usepackage{amsmath}
\usepackage{amssymb}
\usepackage{amsthm}
\usepackage{mathrsfs}
\usepackage{tabularx}
\usepackage{booktabs}
\usepackage{gensymb}
\usepackage{textcomp}

\usepackage{caption}
\usepackage[labelformat=simple]{subcaption}

\DeclareCaptionLabelFormat{subcaptionlabel}{\normalfont(\textbf{#2}\normalfont)}
\begin{document}



\section{Introduction} \label{introduction}


Unmanned aerial vehicles (UAVs) are becoming increasingly popular in the civil and military markets. They play an important role in terrain mapping, environmental data collection, border monitoring, security force assistance, fire monitoring, etc. In most cases, the {UAV} communication systems operate in VHF and UHF bands and use two communication channels to send control commands to the aircraft and information from the aircraft to the base station (BS).

The communication links used by UAVs are subject to the effects of the environment such as fading, which leads to signal degradation and compromised performance and thus impairs communication. Although this degradation can be more easily noted in urban areas, the effects caused by vegetation can also have a significant impact, due to the large quantity of scatters~\cite{ref1-journal}. The signal can go through multiple paths to the reception, spreading the signal in the time domain, which causes selectivity in the frequency domain. As the channel coherence bandwidth becomes lower than the transmitted signal bandwidth, the reception experiences inter-symbolic interference and the bit error rate (BER) floor.

Another important factor that can impair communication is  the Doppler effect caused by the UAV mobility in relation to the BS, which spreads the signal in the frequency domain and, consequently, causes selectivity in the time domain~\cite {ref2-journal}. In this case, the symbol duration is higher than the coherence time of the channel, even at high transmission rates. This causes severe changes in the amplitude and phase of the signal during the symbol duration, increasing BER. 

The characterization of the wireless channels aims to obtain a mathematical model that describes the effects on the received signal more properly. This is important to design solutions to mitigate such effects, especially in UAV communications, that directly affect the flight safety and distance range. In 2013, the United States of America created a special committee (SC-228), formed by the Technical Commission for Radio for Aeronautics (RTCA), whose objective was to define the minimum performance standards for UAV operation in regard to the command and control data link~\cite {ref3-journal}. The first version of the standards document was published in~\cite {ref4-journal}.

One way to characterize the channel is to use first-order statistics such as the cumulative distribution function---CDF---and the probability density function---PDF---to describe the channel. However, an important feature in mobile communication channels is the presence of the Doppler effect, which can cause communication problems as this effect becomes stronger. The communication channels used by these aircraft are subject to this effect, due to the relative mobility between the aircraft and the ground station~\cite{ref5-journal}. Therefore, it is important to characterize the Doppler effect and its impact as the aircraft moves. The characterization of the Doppler effect can be performed by second-order statistical methods, such as the level crossing rate---LCR---which corresponds to the frequency at which the signal strength crosses a certain threshold in the positive or negative direction~\cite{ref6-journal,ref7-journal,ref8-journal}.

More specifically, the communication between the aircraft and the ground station requires high reliability due to the challenges imposed by the environment, such as relief variation, vegetation, and urbanization. In Brazil, an important biome present in the northeast region is the Caatinga, which means “white forest” in the Tupi language. This region comprises an area of 850,000 km$^{2}$ covering 10 Brazilian states. It is a typical tropical vegetation with trees up to 8 m high and covers a population of more than \mbox{53,000,000 inhabitants}, which makes it a significant economic area~\cite {ref9-journal}. The use of UAVs in this region has grown over the years, and there is not much work related to the characterization of radio frequency channels in this environment. Another significant factor that influences the communication signal is the presence of lakes, as the surface of these places may reflect, scatter, or attenuate the electromagnetic signal.

Many studies have been carried out to characterize the communication channel used by UAVs, as can be seen in Table~\ref{tab_related_wroks}. We notice that most of the works focus on large-scale fading statistics, such as the path loss exponent and shadowing, but only few of them characterize the communication channel in relation to the Doppler effect. This paper aims to identify the channel model that best describes the large and small-scale effects in the environments of the Caatinga and lakes. For this, we use the Phantom 3 Standard UAV equipped with an XBee module to transmit data on the frequency channel of 915~MHz. Measurement campaigns were carried out with the UAV flying at different heights and speeds in three regions defined by the environments: (i) lake, (ii) Caatinga biome, and (iii) both lake and Caatinga. From the measured data, we characterize the large-scale attenuation (path loss and shadowing) and small-scale fading, comparing fading statistical distributions and using second-order statistics such as the level crossing rate (LCR) to characterize the Doppler scattering.

The rest of the paper is organized as follows. Section~\ref{materials} describes the  materials used for the wireless channel measurements. We present the mathematical formulation and the measurement data processing in Section~\ref{formulation}. The scenario where the measurements were obtained is depicted in Section~\ref{scenario}. Section~\ref{results} discuss the results of our analysis. Finally,  Section~\ref{conclusion} proceeds with the conclusion of the paper.

\clearpage
\end{paracol}
\nointerlineskip
\begin{specialtable}[H]
\widetable
\small
\caption{\label{tab_related_wroks} Characterization of channels with UAVs.}
\tabcolsep=0.38cm
\begin{tabular}{llllll}
\toprule
\textbf{Work} & \textbf{Frequency} & \textbf{UAV} & \textbf{Scenario} & \textbf{Height} & \begin{tabular}[c]{@{}l@{}}\textbf{Channel}\\ \textbf{Statistics} \end{tabular} \\ \midrule
{\cite{ref10-journal}} & 2 GHz & Airship & Urban & 100--170 m & \begin{tabular}[c]{@{}l@{}}PDF, CDF, AFD,\\ LCR, PSD, AF\end{tabular} \\ \midrule
{\cite{ref11-journal}} & 2 GHz & Airship & Urban & 150--300 m & PL \\ \midrule
{\cite{ref12-journal}} & \begin{tabular}[c]{@{}l@{}}5.76 GHz\\ 1.817 GHz\end{tabular} & Hexacopter & Suburban & 0--50 m & \begin{tabular}[c]{@{}l@{}}PL, SF, K, RMS,\\  CDF\end{tabular} \\ \midrule
{\cite{ref13-journal}} & 4.3 GHz & Quadcopter & \begin{tabular}[c]{@{}l@{}}Open field,\\  Suburban\end{tabular} & 4--16 m & \begin{tabular}[c]{@{}l@{}}PL, SF, $\mu$, $\varepsilon$, PDF, \\ CDF, RMS, BC\end{tabular} \\ \midrule
{\cite{ref14-journal}} & 2.4  GHz & Hexacopter & \begin{tabular}[c]{@{}l@{}}Laboratory,\\ Open air\end{tabular} & 10--40 m & \begin{tabular}[c]{@{}l@{}}PL, PAS, K, \\ PDF\end{tabular} \\ \midrule
{\cite{ref15-journal}} & 802.11a & Quadcopter & Open field & 15--110 m & PL, PAS, CDF \\ \midrule
{\cite{ref16-journal}} & 802.11a & Quadcopter & \begin{tabular}[c]{@{}l@{}}Open field, \\ Field area\end{tabular} & 20--100 m & PL \\ \midrule
{\cite{ref17-journal}} & 802.11a & Fixed Wing & Aerodrome & 46 m & PL \\ \midrule
{\cite{ref18-journal}} & \begin{tabular}[c]{@{}l@{}}802.11a/g,\\ 900 MHz\end{tabular} & Fixed Wing & Aerodrome, Rural & \begin{tabular}[c]{@{}l@{}}46 m, \\ 107--274 m\end{tabular} & PL \\ \midrule
{\cite{ref19-journal}} & GSM, UMTS & \begin{tabular}[c]{@{}l@{}}Fixed Wing, \\ Capture balloon\end{tabular} & Urban, Rural & 0--500 m & PL \\ \midrule
{\cite{ref20-journal}} & \begin{tabular}[c]{@{}l@{}}GSM, UMTS,\\ LTE\end{tabular} & Weather balloon & Urban & 11--18 m & PL \\ \midrule
{\cite{ref21-journal}} & LTE (800 MHz) & Hexacopter & Rural & 15--100 m & PL, SF \\ \midrule
{\cite{ref22-journal}} & LTE (850 MHz) & Quadcopter & Suburban & 15--120 m & PL, SF \\ \midrule
{\cite{ref23-journal}} & 2 GHz & Airship & Urban, Wooded Region & 100--170 m & \begin{tabular}[c]{@{}l@{}}CDF, DG,\\  AFD, LCR\end{tabular} \\ \midrule
\begin{tabular}[c]{@{}l@{}}\cite{ref24-journal}  \\ \cite{ref25-journal}\end{tabular} & 5.8 GHz & Octocopter & Residential & - & \begin{tabular}[c]{@{}l@{}}RMS, DS, \\ CDF\end{tabular} \\ \midrule
{\cite{ref26-journal}} & 802.11b/g & Fixed Wing & Agricultural region & 75 m & AF, DG \\ \midrule
{\cite{ref27-journal}} & \begin{tabular}[c]{@{}l@{}}PCS, AWS, \\ 700 MHz\end{tabular} & Quadcopter & Mix Suburban & 122 m & PL, CDF \\ \midrule
{\cite{ref28-journal}} & \begin{tabular}[c]{@{}l@{}}EDGE, HSPA+,\\ LTE\end{tabular} & Hexacopter & - & 10--100 m & RTT, J \\ \midrule
{\cite{ref29-journal}} & 909 MHz & Quadcopter & \begin{tabular}[c]{@{}l@{}}Open field,\\  Simulated Village\end{tabular} & 40--60 m & PL, PES \\ \midrule
{\cite{ref30-journal}} & 2/3.5/5.5 GHz & HAP airship & \begin{tabular}[c]{@{}l@{}}Built-up areas\\  \end{tabular} & - & SF \\ \midrule
{\cite{ref31-journal}} & 2.585 GHz & Hexacopter & \begin{tabular}[c]{@{}l@{}}Suburban\\ \end{tabular} & 15--300 m & PDP, RMS, DS \\ \midrule
{\cite{ref32-journal}} & 3.4/3.8 GHz & Commercial UAV & \begin{tabular}[c]{@{}l@{}}Open area\\ \end{tabular} & 5--15 m & PDP, RMS \\ \midrule
{This work} & 915~MHz & Quadcopter & \begin{tabular}[c]{@{}l@{}}Lake,\\  Caatinga\end{tabular} & 8--80 m & \begin{tabular}[c]{@{}l@{}}PL, LCR, CDF, DS, SF,\\  K, $\mu$, $\varepsilon$, $m$, $\beta$, $\sigma$\\ 
\end{tabular}\\\bottomrule
   \end{tabular}

 \begin{tabular}{c}
\multicolumn{1}{p{\linewidth-1.5cm}}{\footnotesize \justifyorcenter{AF: {{Autocorrelation function}
}, AFD: {Average fade duration}, BC: {Coherence bandwidth}; CDF: {Cumulative distribution function}, DG: {Diversity gain}, DS: {Doppler spread}, J: {Jitter}, K: {Rician factor}; LCR: {Level crossing rate}, PAS: {Power azimuth spectrum}, PDF: {Probability density function}; PDP: {Power delay profile}, PES: {Power elevation spectrum}, PL: {Path loss}, RMS: RMS {delay} spread; RTT: {Round trip time}, SF: {Shadow fading}, e $\mu$, $\varepsilon$: {Mean and standard deviation of Nakagami} $m$ {factor}; $\beta$: {Weibull shape parameter}, $\sigma$: {Rayleigh parameter}.}}
\end{tabular}

\end{specialtable}

\begin{figure}[H]
\widefigure
\includegraphics[width=0.9\textwidth]{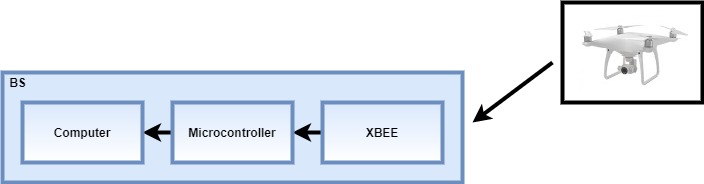}
\caption{\label{esquema_plataforma} Scheme of reception and storage of the wireless signal measurement data.}
\end{figure}

\begin{figure}[H]
\widefigure
\begin{subfigure}{.49\textwidth}
  \includegraphics[width=0.939\linewidth]{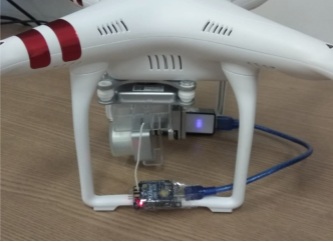}  
  \caption{}
  \label{plataforma_a}
\end{subfigure}~
\begin{subfigure}{.49\textwidth}
  \includegraphics[width=1.0\linewidth]{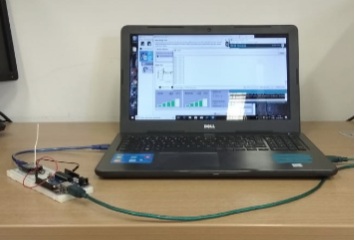}  
  \caption{}
  \label{plataforma_b}
\end{subfigure}

\caption{Equipment used in measurement campaigns. (\textbf{a}) UAV with XBee and power battery modules. (\textbf{b}) Base station for collecting measurement data.}
\label{fig:fgig}
\end{figure}

\begin{paracol}{2}
\switchcolumn


\vspace{-16pt}
\section{Materials} \label{materials}

The equipment used in the measurements campaign was composed of a UAV and XBee modules for wireless communications.

\subsection{Unmanned Aerial Vehicle}


The UAV used in the measurement campaign is the DJI Phantom 3 Standard model manufactured by the Chinese company DJI. This aircraft has an autonomous flight mode, making it possible to insert waypoints in a defined route. Its range is restricted to the battery autonomy time, which is approximately 25 min. Another way to fly is to use the manual mode, in which case the aircraft operator uses a radio control to send the commands to the aircraft. This flight mode allows reaching a maximum range of 1000~m.

The aircraft has a mass of 1216~g, ascending speed of 5~m/s, descending speed of 3~m/s, diagonal length without flight pallets of 350~mm, maximum speed of 16~m/s, and built-in GPS with a maximum error of 0.5~m (vertical) and 1.5~m (horizontal). The measurement campaign used the autonomous flight mode, as it allows for defining the starting and arrival point and completing the route at a chosen speed~\cite{ref33-journal}.

\subsection{XBee Module}

The XBee 900HP PRO S3 module is a platform that offers wireless connectivity for point-to-point communication or in mesh networks with up to 128 nodes. The device consists of transmitters with configurable power between 10 and 250~mW and an omnidirectional antenna with a gain of 2~dbi. It has a variable transmission rate between 10 and 200~kbps. Its receiver has a minimum sensitivity of $-$110~dBm, which allows a maximum communication range between modules of up to 6.5~km~\cite {ref34-journal}.

In order to obtain wireless channel measurements, we shipped the XBee device in the UAV and configured  the following parameters: unicast communication with IEEE 802.15.4 protocol (ZigBee)~\cite {ref35-journal}, transmission power of 250~mW, and interval between sending packets of 300~ms. The UAV went up to the programmed height (8 or 80~m) depending on the specifications of the measurements. During the flight to a specific point without changing the height and with constant speed, the embedded device sends the packages to the BS that measures the received signal.

The base station consists of a notebook with Intel (R) Core (TM) i5-7200U 2.70 GHz processor, 8 GHz RAM, and 64-bit Windows 10 Home Single Language operating system; a microcontroller; and the XBee module configured to receive the packets, as presented in Figure~\ref{esquema_plataforma}. When the signal arrives at the XBee module, the microcontroller is configured to capture the signal strength of the received packet and send the data to be saved in the computer.

\end{paracol}
\nointerlineskip
\begin{paracol}{2}
\switchcolumn

Figure~\ref{fig:fgig}a shows the UAV with the XBee module responsible for sending data. Figure~\ref{fig:fgig}b shows the base station composed of the XBee module, the microcontroller, and the computer to power the system and save the data.

\clearpage
\end{paracol}
\nointerlineskip
\begin{paracol}{2}
\switchcolumn


\vspace{-10pt}
\section{Mathematical Formulation and Methodology} \label{formulation}

Several procedures are carried out to process the data collected from the measurement campaigns. In order to keep the quality of the collected data higher, the measurement campaigns that had a success rate of received packages less than 98\% were discarded due to the low SNR. Such a threshold was obtained empirically in previous experiments, whose results presented low adherence when comparing experimental and theoretical distribution functions.

\subsection{Filtering}\label{filtragem}

The first step after collecting the measured signal power level data is to separate small-scale fading from large-scale attenuation. For this, we used the moving average filter given by~\cite{ref36-journal}.

\begin{equation}
\label{eq_filter}
    y[i]=\frac{1}{N+M+1}\sum_{K=-N}^{M}x[i+k],
\end{equation}

\noindent where ${x [i + k]}$ are the signal samples, {$i$ is} the filter sample index, ${y [i]}$ is the $i$-th filter output sample, and $N$ and $M$ are the number of samples of the input signal before and after the $i$-th sample, respectively. A symmetric window was considered, i.e., $N = M$.

The size of the moving average window is obtained empirically, i.e., we choose the smallest possible window whose output samples pass on the Kolmogorov---Smirnov (KS) test~\cite{ref37-journal} for any of the fading distributions (Rayleigh, Rice, Nakagami, or Weibull) with a 95\% confidence index. In this way, each measurement campaign has its specific filtering window.

\subsection{Large-Scale Attenuation}

As the small-scale fading samples pass successfully on the KS test, we characterize the large-scale attenuation in order to determine the parameters for path loss and log-normal shadowing. The large-scale attenuation can be modeled as~\cite{ref40-journal}


\begin{equation}
    \label{sinal_recebido_eq}
    \overline{PL}(dB)=\overline{PL}(d_0)+10n\log\left( \frac{d}{d_0} \right)+X_\sigma,
\end{equation}

\noindent where $d$ is the distance between transmitter and receiver, $n$ is the exponent of the path loss, $\overline{PL} (d_0)$ the attenuation in dB at the start point, $d_0$ the start point of measurement, and $X_\sigma$ is a zero-mean normal random variable with standard deviation~$\sigma$ that models the shadowing effect~\cite{ref41-journal, ref42-journal}. The value of $n$ is obtained by the linear regression method~\cite{ref38-journal, ref39-journal}.



\subsection{Small-Scale Fading}

We used the maximum likelihood estimation (MLE) method to estimate the parameters of the Rayleigh, Rice, Nakagami, and Weibull distributions~\cite{ref43-journal}. Then, the theoretical cumulative probability functions (CDFs) were generated and compared with the empirical CDFs obtained from the small-scale fading samples. Henceforth, it is possible to verify which distribution properly describes the channel based on the KS test with a confidence level equal to or greater than 95\%. This procedure served to define the size of the filtering window.

\subsubsection{Level Crossing Rate}

From the selected fading distribution, it is possible to calculate its theoretical level crossing rate, in order to estimate the Doppler spread. The theoretical functions of the LCR for the Rayleigh, Rice, Nakagami, and Weibull are respectively~\cite{ref44-journal,ref45-journal}

\begin{equation}
\label{eq_lcr_rayleigh}
    N_{Rayleigh}(\rho) = \sqrt{2\pi}f_d \; \rho \; \exp(-\rho^2),   
\end{equation}


\begin{equation}
\label{eq_lcr_rice}
    N_{Rice}(\rho) = \sqrt{2\pi(k+1)}f_d \; \rho \; \exp(-K -(k+1)\rho^2) \; I_0(2\sqrt{K(K+1))}\rho,
\end{equation}



\begin{equation}
\label{eq_lcr_nakagami}
    N_{Nakagami}(\rho) = \sqrt{2\pi}f_d\; \frac{m^{m-1/2}}{\Gamma(m)} \rho^{2m-1}\exp(-m\rho^2). 
\end{equation}

\noindent and

\begin{equation}
\label{eq_lcr_weibull}
    N_{Weibull}(\rho) = \sqrt{2\pi}f_d  \left ( \frac{\rho}{\sqrt(a)} \right )^{\beta/2} \; \exp\left [ -\left ( \frac{\rho}{\sqrt(a)} \right )^\beta \right ],
\end{equation}

\noindent where $r$ represents the fade samples, $f_d$ is the {Doppler frequency}, $\rho = r/ \sqrt{\Omega}$,  $\Omega = E[r^2]$, $I_0$ is the modified Bessel function of the first kind and order $0$, and $\Gamma(.)$ is the Gamma function. The Rician parameter $K$ represents the ratio between the power of the dominant signal (\mbox{in line} of sight) and the power of the multipath components. The Nakagami parameter $m$ refers to the quantity of multipath clusters, while the Weibull parameter $\beta$ is related to the non-linearity of the channel, and $a = 1/\Gamma(1 + (\beta/2))$.

\subsubsection{{Doppler Frequency}}
\label{sub_sec_doppler}

The LCR function is written as a function of the Doppler frequency. Thus, with both theoretical and empirical the values of LCR, it is possible to estimate the Doppler frequency $f_d$~\cite{ref46-journal} as

\begin{equation}
    \hat{f}_d = \frac{LCR_{Empirical}}{LCR_{Theoretical\_normalized}},
    \label{eq_fm_tcn}
\end{equation}


\noindent where  $LCR_{Theoretical\_normalized}$ can be obtained from Equations \eqref{eq_lcr_rayleigh}--\eqref{eq_lcr_weibull} with $f_d = 1$.


Then, the relative speed between the transmitter and receiver can be estimated as $\hat{v} =\lambda \cdot \hat{f}_d/\;\cos(\theta)$, where $\lambda$ is the wavelength of the carrier, and $\theta$ refers to the angle of the horizontal axis of the antenna to the BS. It is important to make it clear that these equations refer to the calculation of second-order statistics for a given power level. Thus, for each calculated level, there will be a different value for $\hat{f}_d$.


\section{Measurement Scenario}\label{scenario}


Measurement campaigns were carried out in Brazil in the city of Macaíba in the state of Rio Grande do Norte. During the campaign, we verified the influence of the wind on the trajectory and on the average speed of the aircraft at low speeds. Thus, the speed determined in the aircraft's control software changed during the flight, since all flights were at low speed, and gusts of wind were strong enough to modify the instantaneous speed and positioning of the UAV. Table~\ref{tab_clima} shows the wind speed at the time of the campaigns, as well as the local temperature, both obtained from a local weather station.

\begin{specialtable}[H]
\caption{\label{tab_clima} Table of temperature and average wind speed on the day of measurement.}
\tabcolsep=1.15cm
\begin{tabular}{lll}
\toprule 
\textbf{Schedule} & \textbf{Temperature} & \textbf{Average Wind Speed}
\\\midrule 
10:00 & 31 $\degree$C & 31~km/h \\
11:00 & 31 $\degree$C & 28~km/h \\
12:00 & 31 $\degree$C & 33~km/h \\
13:00 & 31 $\degree$C & 28~km/h \\
14:00 & 30 $\degree$C & 33~km/h \\
15:00 & 30 $\degree$C & 35~km/h \\
16:00 & 30 $\degree$C & 28~km/h \\
17:00 & 29 $\degree$C & 24~km/h \\
\bottomrule 
\end{tabular}
\end{specialtable}

\subsection{Measurement Scenario 1: Lake}

The measurement scenario 1 corresponds to flights over a lake of approximately 12,000 square meters and a maximum depth of 12 m. The UAV was configured to perform straight-line flights from a 5~m point horizontally from the BS. The height and speed of measurements can be seen in Table~\ref{tab_campanha_lago}. 

Figure~\ref{fig_lago} shows images taken from the UAV over the lake at the time of flight at different heights. Figure~\ref{fig_lago}c presents the direction of the flight. The wind was blowing on the side of the aircraft  and in the opposite direction of the flight. 

\begin{specialtable}[H]
\caption{\label{tab_campanha_lago} Scenario~1 parameters.}
\tabcolsep=1.4cm
\begin{tabular}{lll}
\toprule 
\textbf{Velocity} & \textbf{Height} & \textbf{Traveled Distance} \\\midrule 
1~km/h & 8~m & 120~m \\
1~km/h & 80~m & 150~m\\
3~km/h & 8~m & 120~m \\
3~km/h & 80~m & 150~m\\
\bottomrule 
\end{tabular}
\end{specialtable}


\end{paracol}
\nointerlineskip
\begin{figure}[H]
\widetable
  \begin{subfigure}{0.31\textwidth}
    \includegraphics[width=\linewidth]{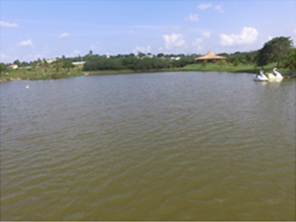}
    \vspace{2pt}
    \caption{} 
    \label{lago1}
  \end{subfigure}%
  \hspace*{\fill}   
  \begin{subfigure}{0.31\textwidth}
    \includegraphics[width=\linewidth]{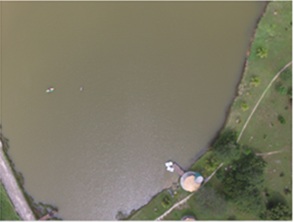}
        \vspace{2pt}
    \caption{} 
    \label{lago2}
  \end{subfigure}%
  \hspace*{\fill}   
  \begin{subfigure}{0.31\textwidth}
    \includegraphics[width=\linewidth]{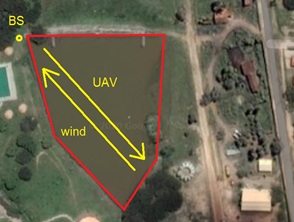}
        \vspace{2pt}
    \caption{}
    \label{lago3}
  \end{subfigure}
\\
\vspace{2pt}
\caption{Lake photos taken at different heights. (\textbf{a}) Photo taken with UAV over the lake at a height of 8~m. (\textbf{b}) Photo taken with UAV over the lake at a height of 80~m. (\textbf{c}) Showing the base station (BS) and the lake area.} 
\label{fig_lago}
\end{figure}
\begin{paracol}{2}
\switchcolumn

\vspace{-20pt}
\subsection{Measurement Scenario 2: Caatinga}


Scenario 2 corresponds to the Caatinga vegetation, covering an area of approximately 37,000 square meters. Two flights over the vegetation were carried out, both at a height of 80~m and speeds of 1 and 3~km/h, as exemplified in Table~\ref{tab_regiao_II}. In these flights, the aircraft suffered wind force in the same direction as the flight and laterally. In this windy situation, the aircraft had to travel different distances for safety reasons, as the strong wind could take the aircraft out of the planned trajectory, and we could lose contact with the aircraft. 
The flight start point was 5~m horizontally from the BS, and the aircraft was programmed to fly in a straight line.

\begin{specialtable}[H]

\caption{\label{tab_regiao_II} Scenario~2 parameters.}
\tabcolsep=1.38cm
\begin{tabular}{lll}
\toprule 
\textbf{Velocity} & \textbf{Height} & \textbf{Traveled Distance }
\\\midrule 
1~km/h & 80~m & 130~m \\
3~km/h & 80~m & 250~m \\\bottomrule 
\end{tabular}
\end{specialtable}

Figure~\ref{fig_caatinga} shows the total aerial view of scenario 2 and a photo taken by the UAV at an 80-meter height over the region.

\end{paracol}
\nointerlineskip
\begin{figure}[H]
\widefigure
\begin{subfigure}{.5\textwidth}
  
  \includegraphics[width=1\linewidth]{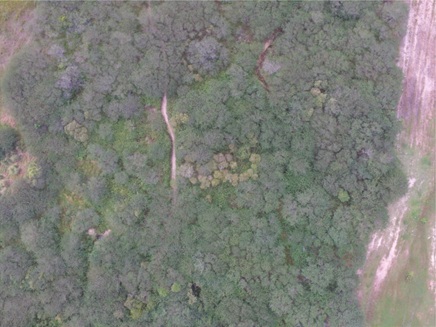}  
  \caption{}
  \label{caatinga_a}
\end{subfigure}~
\begin{subfigure}{0.27\textwidth}
  
  \includegraphics[width=1\linewidth]{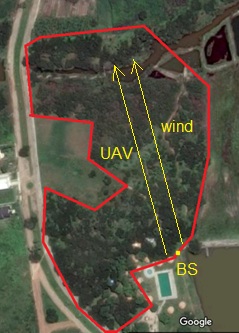}  
  \caption{}
  \label{caatinga_b}
\end{subfigure}

\caption{Caatinga biome region. (\textbf{a}) Flight region over the Caatinga vegetation. (\textbf{b}) Area of the flight over the Caatinga vegetation.}
\label{fig_caatinga}
\end{figure}
\begin{paracol}{2}
\switchcolumn

\vspace{-14pt}
\subsection{Measurement Scenario 3: Mixed}

Region III corresponds to an area tangent to the lake, as can be seen in Figure~\ref{fig_misto}, and which includes a small area of the Caatinga. The flights were performed at a height of 80~m and a speed of 1 and 3~km/h, as shown in Table~\ref{Misto_tabela}. The aircraft started its flight 5~m distant horizontally from the BS.

\begin{specialtable}[H]

\caption{\label{Misto_tabela} Scenario~3 parameters.}
\tabcolsep=1.4cm
\begin{tabular}{lll}
\toprule 
\textbf{Velocity} & \textbf{Height} & \textbf{Traveled Distance}
\\\midrule 
1~km/h & 80~m & 150~m \\
3~km/h & 80~m & 150~m \\
\bottomrule 
\end{tabular}
\end{specialtable}

Figure~\ref{fig_misto} shows the region tangent to the lake and an image captured by the UAV when flying over the small area of the Caatinga at a height of 80~m.

\end{paracol}
\clearpage
\nointerlineskip
\begin{figure}[H]
\widefigure
\begin{subfigure}{.5\textwidth}
  
  \includegraphics[width=0.9\linewidth]{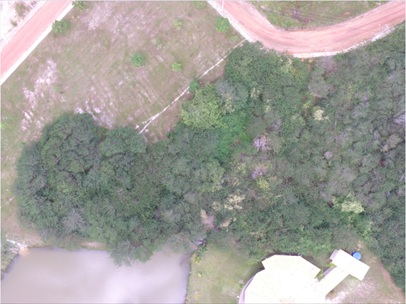}  
  \caption{}
  \label{mistto_a}
\end{subfigure}~\begin{subfigure}{0.3\textwidth}
  
  \includegraphics[width=1\linewidth]{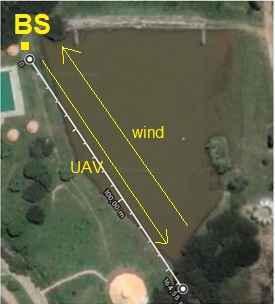}  
  \caption{}
  \label{misto_b}
\end{subfigure}
\vspace{4pt}
\caption{Mixed scenario. (\textbf{a}) Mixed scenario photo taken by the UAV at 80~m. (\textbf{b}) Full view of the mixed scenario.}
\label{fig_misto}
\end{figure}

\begin{figure}[H]
\widefigure
\begin{subfigure}{.5\textwidth}
  
  \includegraphics[width=1.0\linewidth]{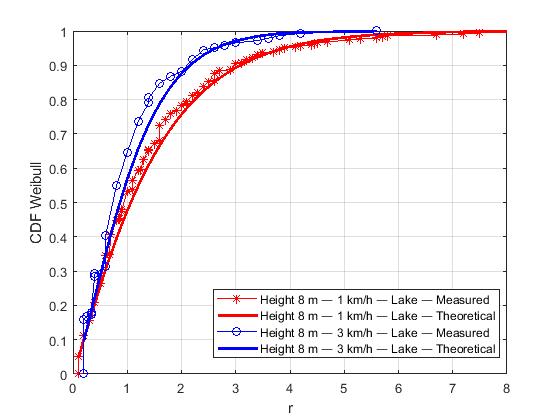}  
  \caption{}
  \label{cdf_lago_a}
\end{subfigure}~
\begin{subfigure}{.5\textwidth}
  
  \includegraphics[width=1.0\linewidth]{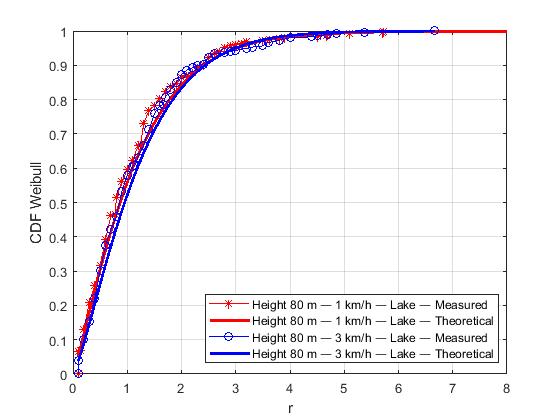}  
  \caption{}
  \label{cdf_lago_b}
\end{subfigure}

\caption{Estimated CDFs over the lake for the same height. (\textbf{a}) CDFs over the lake at 8~m. (\textbf{b}) CDFs over the lake at 80~m.}
\label{cdf_lago}
\end{figure}

\begin{figure}[H]
\widefigure
\begin{subfigure}{.5\textwidth}
  
  \includegraphics[width=1.0\linewidth]{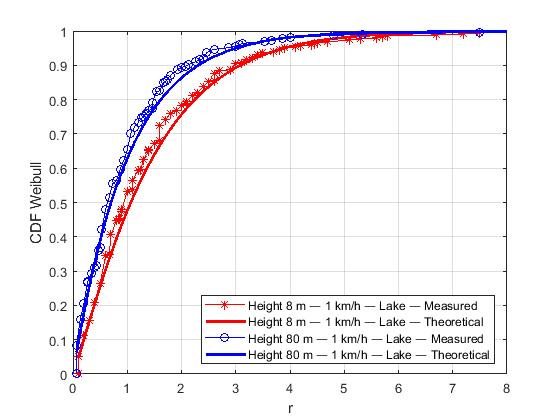}  
  \caption{}
  \label{cdf_lago_2_a}
\end{subfigure}~
\begin{subfigure}{.5\textwidth}
  
  \includegraphics[width=1.0\linewidth]{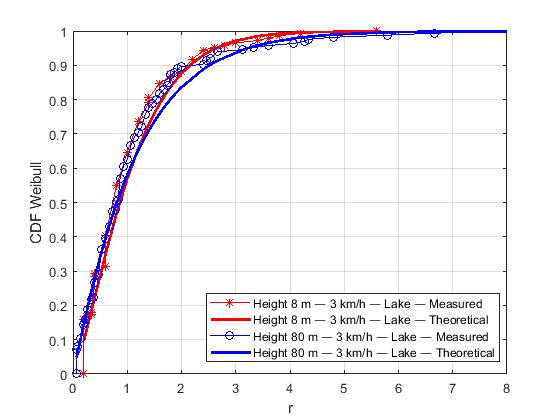}  
  \caption{}
  \label{cdf_lago_2_b}
\end{subfigure}

\caption{Estimation of CDFs over the lake at the same speed. (\textbf{a}) CDFs over the lake at 1~km/h. (\textbf{b}) CDFs over the lake at 3~km/h.}
\label{cdf_lago_2}
\end{figure}

\begin{figure}[H]
\widefigure
\begin{subfigure}{.5\textwidth}
  
  \includegraphics[width=1.0\linewidth]{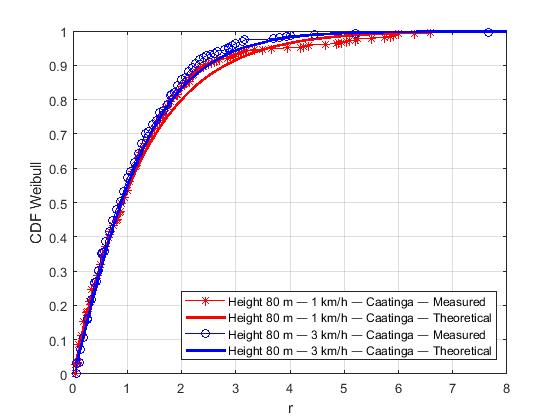}  
  \caption{}
  \label{cdf_caatinga}
\end{subfigure}~
\begin{subfigure}{.5\textwidth}
  
  \includegraphics[width=1.0\linewidth]{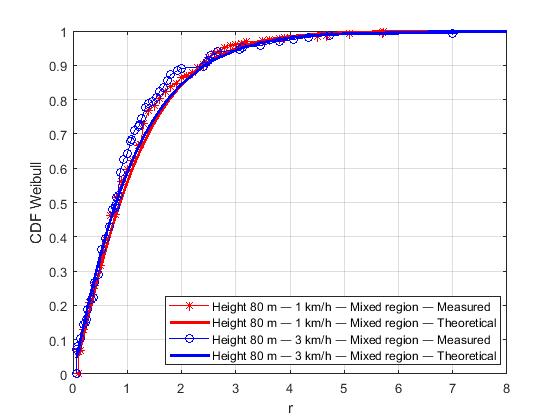}  
  \caption{}
  \label{cdf_misto}
\end{subfigure}

\caption{Estimation of CDFs for the Caatinga and Mixed environments. (\textbf{a}) CDFs for the Caatinga region. (\textbf{b}) CDFs for mixed region.}
\label{cdf_caatinga_misto}
\end{figure}

\begin{paracol}{2}
\switchcolumn


\vspace{-8pt}
\section{Results}\label{results}

After filtering the received signal, the linear regression method was used in the fading samples to estimate the path loss. Figure~\ref{path_loss_lago} shows the path loss for flights over the lake (scenario 1). For higher altitude flights, the signal presented a positive exponent (gain) and negative exponent for the route of lower altitudes. This effect is related to the first partially cleared Fresnel zone~\cite{ref5-journal}, as also observed in~\cite{ref12-journal}. 
In flights at greater heights, the path loss decreases as distance increases, as observed in~\cite{ref22-journal, ref47-journal}. {In~\cite{ref47-journal}, an experimental path loss study was carried out with the UAV flying at an altitude of 1~km. The work showed that the power received at the base station increases as the aircraft moves away from the base station, until it reaches its maximum value and then begins its decline, that is, the intensity of power received at the base station decreases as the aircraft moves away. This result of gain with increasing distance was observed in this work for all environments, except for flights at 8 m of altitude, as observed in Figure~\ref{path_loss_lago}. In~\cite{ref49-journal}, {the work analyzed} 
 the path loss of a mobile communication system  in a forest, and the results showed that the values varied in a range between 1.6 and 4.4. The same result was observed in this work, as shown in Table~\ref{tab_expoente}. The exceptions were for the low altitude flight environment over the lake, which presented a high exponent modulus of path loss.}


\begin{figure}[H]
\includegraphics[width=0.55\textwidth]{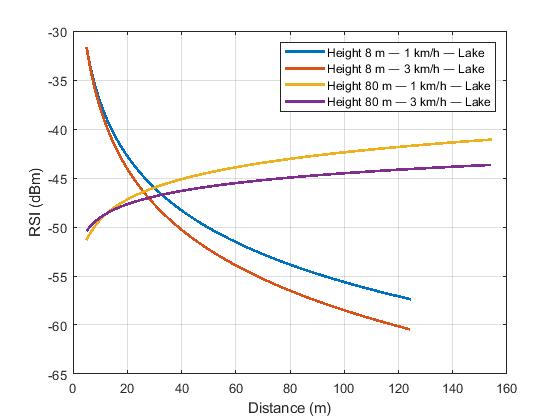}
\caption{\label{path_loss_lago} {Path} 
 loss of flights over the lake.}
\end{figure}

Figure~\ref{path_loss_mata} shows the path loss to the Caatinga scenario with the flight at an 80-meter height and speeds of 1 and 3~km/h. The same behavior of Figure~\ref{path_loss_lago} is observed considering the flight at a height of 80~m, 
with higher received power levels as distance increases.

\begin{figure}[H]

\includegraphics[width=0.55\textwidth]{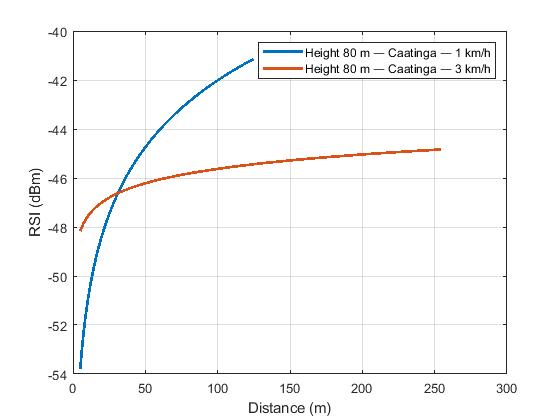}
\caption{\label{path_loss_mata} {Path} loss of flights over the Caatinga.}
\end{figure}


The scenario tangent to the lake presented the same trend, as shown in Figure~\ref{path_loss_misto}. Therefore, the communication between the UAV and the BS at the moment the aircraft flies over the BS may be hampered due to the use of dipole antennas. Appendix~\ref{ap1} illustrates the dipole antenna radiation pattern, showing the decreasing antenna gain for upper directions (Figure~\ref{antenna_drones}). However, when the UAV starts to move away from the base station, the antenna gain starts to increase due to the change in the main Fresnel zone.

\begin{figure}[H]

\includegraphics[width=0.55\textwidth]{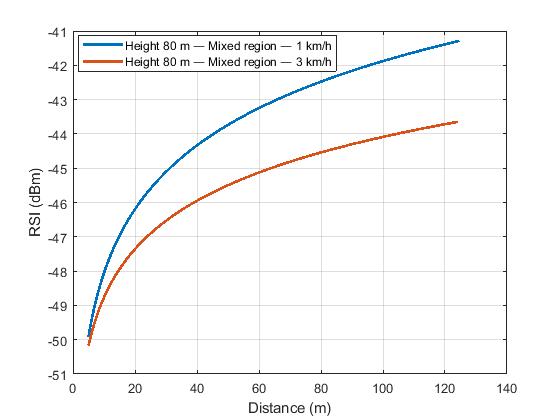}
\caption{\label{path_loss_misto} {Path} loss of flights over the mixed region.}
\end{figure}

Equation~(\ref{sinal_recebido_eq}) was used to estimate the path loss exponent, as shown in Table~\ref{tab_expoente}. As expected, path loss exponents for low-altitude flights have a negative exponent. For the purpose of comparison, reference values for path loss exponents are 2 for free space and 4 to 6 for channels obstructed by buildings. Thus, low-altitude flights have a more severe path loss.

\begin{specialtable}[H]

\caption{\label{tab_expoente}Path loss exponent for different environments.}
\tabcolsep=0.55cm
\begin{tabular}{llll}
\toprule 
\textbf{Environment} & \textbf{Path Loss Exponent} & \textbf{Speed (km / h)} & \textbf{Height (m)} \\\midrule 
Lake & $-$7.8 & 1 & 8 \\
Lake & $-$8.9 & 3 & 8 \\
Lake & 2.9 & 1 & 80 \\
Lake & 2.0 & 3 & 80 \\
Mixed region & 3.7 & 1 & 80 \\
Mixed region & 3.8 & 3 & 80 \\
Caatinga & 3.7 & 1 & 80 \\
Caatinga & 1.9 & 3 & 80 \\
\bottomrule 
\end{tabular}
\end{specialtable}

Subtraction of the estimated path loss from large-scale fading samples was performed to estimate the log-normal shadowing. Table~\ref{tab_sombreamennto} shows the values of mean and standard deviation of log-normal shadowing for some sizes of the filtering window to obtain the fading. The average is expected to approach zero for good adherence to the measured results. The highest values of the standard deviation are observed with the flights over the lake with the lowest height. These scenarios also present higher values of the path loss exponent.

\end{paracol}
\nointerlineskip
\begin{specialtable}[H]
\widetable
\caption{\label{tab_sombreamennto} Shadowing parameters for different environments.}
\tabcolsep=0.44cm
\begin{tabular}{llllll}
\toprule 
\textbf{Environment} & \textbf{Height (m)} & \textbf{Speed (km/h) }& \textbf{Average (\boldmath{$\mu$})} & \textbf{Standard Deviation ($\sigma$)} & \textbf{Window} \\\midrule 
Lake & 8 & 1 & $-$0.036457 & 4.9594 & 10 \\
Lake & 8 & 3 & $-$0.00010684 & 5.1219 & 5 \\
Lake & 80 & 1 & 0.014171 & 1.5940 & 15 \\
Lake & 80 & 3 & 0.019407 & 1.3563 & 15 \\
Mixed region & 80 & 1 & 0.021741 & 1.8036 & 10 \\
Mixed region & 80 & 3 & 0.023567 & 1.4659 & 20 \\
Caatinga & 80 & 1 & 0.0089469 & 2.6579 & 10 \\
Caatinga & 80 & 3 & $-$0.0052443 & 3.0010 & 15 \\
\bottomrule 
\end{tabular}
\end{specialtable}
\begin{paracol}{2}
\switchcolumn


We use the maximum likelihood method to estimate the parameters of the Nakagami, Rice, Rayleigh, and Weibull distributions, as shown in Table~\ref{tab_parametros}. These results display the parameters of distributions from the measured samples.

The Kolmogorov--Smirnov (KS) test was used to compute the similarity of the empirical CDF (from measurements) and theoretical CDFs of all distributions at a significance rate of 95\%. The KS test indicates Weibull as the best-fit distribution  for the collected data. Using the estimated parameters $\beta$ of the Weibull distribution (Table~\ref{tab_parametros}), the theoretical CDFs were computed and plotted  to compare with the empirical CDFs, as shown in Figures~\ref{cdf_lago}--\ref{cdf_caatinga_misto}. We can notice the good curve-fitting of empirical and estimated CDFs. 

Figure~\ref{cdf_lago} presents the results for the Lake region for a height and different speeds in the same plot. We observe a higher small-scale fading variation at a lower speed (Figure~\ref {cdf_lago}a). We can deduce that the lower the speed of the aircraft, the worse the flight stabilization, which causes a higher variation in the small-scale fading. Additionally, we can also observe that the results of flights at 80 m do not show significant differences, regardless of the speed (Figure~\ref{cdf_lago}b).

\clearpage
\end{paracol}
\nointerlineskip
\begin{specialtable}[H]
\widetable
\caption{\label{tab_parametros} Parameters of small-scale fading distributions for different environments.}
\tabcolsep=0.4cm
\begin{tabular}{llllllll}
\toprule 
\textbf{Window} & \begin{tabular}[c]{@{}l@{}}\textbf{Nakagami} \\ \textbf{(m, \boldmath{$\Omega$})}\end{tabular} & \textbf{Rice (K)} & \begin{tabular}[c]{@{}l@{}}\textbf{Rayleigh} \\ \textbf{(\boldmath{$\sigma$})}\end{tabular} & \begin{tabular}[c]{@{}l@{}}\textbf{Weibull} \\ \textbf{(\boldmath{$\beta$}, \boldmath{$\lambda$})}\end{tabular} & \begin{tabular}[c]{@{}l@{}}\textbf{Height} \\ \textbf{(m)}\end{tabular} & \textbf{Environment} & \begin{tabular}[c]{@{}l@{}}\textbf{Velocity} \\ \textbf{(km/h)}\end{tabular} \\ \midrule 
10 & \begin{tabular}[c]{@{}l@{}}0.42621,\\ 4.0043\end{tabular} & 0.00033815 & 1.415 & \begin{tabular}[c]{@{}l@{}}1.1251,\\ 1.4969\end{tabular} & 8 & Lake & 1 \\ 
5 & \begin{tabular}[c]{@{}l@{}}0.54171, \\ 1.8553\end{tabular} & 0.00032789 & 0.96315 & \begin{tabular}[c]{@{}l@{}}1.3129, \\ 1.1415\end{tabular} & 8 & Lake & 3 \\ 
15 & \begin{tabular}[c]{@{}l@{}}0.4374,  \\ 2.6966\end{tabular} & 0.00027749 & 1.1612 & \begin{tabular}[c]{@{}l@{}}1.1512, \\ 1.2552\end{tabular} & 80 & Caatinga & 1 \\ 
15 & \begin{tabular}[c]{@{}l@{}}0.43961,\\ 2.6205\end{tabular} & 0.00027911 & 1.1447 & \begin{tabular}[c]{@{}l@{}}1.1688, \\ 1.2184\end{tabular} & 80 & Caatinga & 3 \\ 
10 & \begin{tabular}[c]{@{}l@{}}0.44587, \\ 2.4439\end{tabular} & 0.0002916 & 1.1054 & \begin{tabular}[c]{@{}l@{}}1.1751, \\ 1.1859\end{tabular} & 80 & Mixed region & 1 \\ 
20 & \begin{tabular}[c]{@{}l@{}}0.37927,\\ 3.2004\end{tabular} & 0.00020356 & 1.265 & \begin{tabular}[c]{@{}l@{}}1.0619, \\ 1.2123\end{tabular} & 80 & Mixed region & 3 \\ 
10 & \begin{tabular}[c]{@{}l@{}}0.35111,\\ 2.3994\end{tabular} & 0.00027483 & 1.0953 & \begin{tabular}[c]{@{}l@{}}1.0072, \\ 1.0094\end{tabular} & 80 & Lake & 1 \\ 
15 & \begin{tabular}[c]{@{}l@{}}0.37418, \\ 3.3069\end{tabular} & 0.00019419 & 1.2859 & \begin{tabular}[c]{@{}l@{}}1.0548, \\ 1.2181\end{tabular} & 80 & Lake & 3\\
\bottomrule
\end{tabular}
\end{specialtable}
\begin{paracol}{2}
\switchcolumn


\end{paracol}
\nointerlineskip
\vspace{-14pt}
\begin{paracol}{2}
\switchcolumn



In Figure~\ref{cdf_lago_2}, we organize the results for a speed and different heights in the same plot. At a low speed, lower height flights (Figure~\ref {cdf_lago_2}a) present a higher variation in the small-scale fading.  We speculate that this is related to the drone's inability to remain stable during flights at a low height when it receives strong gusts of wind. On the other hand (Figure~\ref{cdf_lago_2}b), there is no significant difference in the small-scale fading for flights at 3~km/h and different heights.

\clearpage
\end{paracol}
\nointerlineskip
\begin{paracol}{2}
\switchcolumn



Figure~\ref{cdf_caatinga_misto} presents CDFs for the Caatinga (Figure~\ref{cdf_caatinga_misto}a) and Mixed regions (Figure~\ref{cdf_caatinga_misto}b). Similar to the results of the Lake region at 80~m, there is no variation in small-scale fading at higher heights, regardless of the drone's speed. 

\end{paracol}
\nointerlineskip
\begin{paracol}{2}
\switchcolumn


Small-scale fading samples were used to estimate the Doppler frequency. First, the theoretical LCR curves for the Weibull distribution were generated with \mbox{Equation~(\ref{eq_lcr_weibull})}. Then, the LCR values were generated with the small-scale fading samples. Finally, the Equation~(\ref{eq_fm_tcn}) was used to estimate the Doppler.

Figure~\ref{doppler} shows the results of the Doppler frequency estimation. The results for flights on the lake are shown in Figure~\ref{doppler}a. Doppler estimation is similar for almost all flights, except for the one at the height of 8~m and speed of 3~km/h. In this case, the drone was able to develop a more constant speed, since the lake has a natural barrier of vegetation on one of the edges that reduces the action of the wind. In this specific case, the aircraft was able to develop a higher flight speed compared to other measurements {and, consequently, a higher Doppler frequency, as shown in Figure~\ref{doppler}a}. Results for the Caatinga and Mixed regions are presented in Figure~\ref{doppler}b. In these environments, there is no significant difference in the estimated Doppler frequency for different speeds. The aircraft suffered from wind action throughout the course so that there was no control over the average drone speed.

\end{paracol}
\nointerlineskip
\vspace{-12pt}
\begin{figure}[H]
\widefigure
\begin{subfigure}{.5\textwidth}
  
  \includegraphics[width=1.0\linewidth]{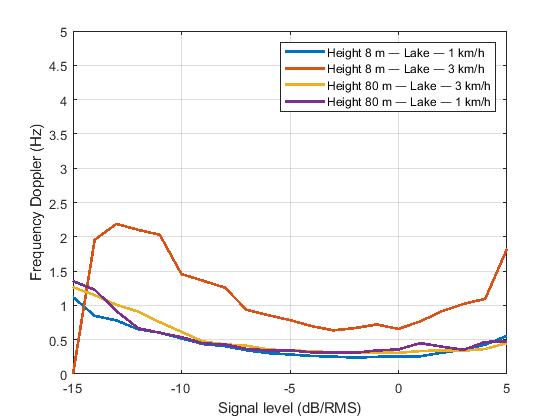}  
  \caption{}
  \label{doppler_a}
\end{subfigure}~
\begin{subfigure}{.5\textwidth}
  
  \includegraphics[width=1.0\linewidth]{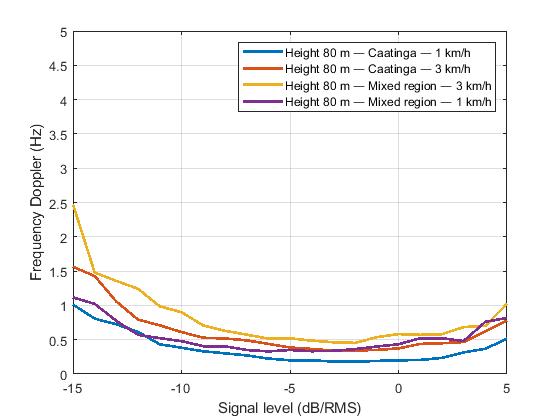}  
  \caption{}
  \label{doppler_b}
\end{subfigure}

\caption{{Doppler frequency} estimation. (\textbf{a}) Doppler frequency for Lake region. (\textbf{b}) Doppler frequency for Caatinga and Mixed regions.}
\label{doppler}
\end{figure}
\begin{paracol}{2}
\switchcolumn



All measurement campaigns were affected by gusts of crosswinds. For instance, in the Mixed environment and over the Lake, the wind is in the opposite direction of the aircraft's flight. We had the inverse situation in the Caatinga region. As the theoretical Doppler frequency depends on the angle between the aircraft and the BS and the drone speed, its value varies even if we consider a unique drone speed. Table~\ref{Variacao_doppler} shows the range (minimum and maximum values) of the theoretical Doppler frequency. Comparing these theoretical values to the estimated ones from the measurements (Figure~\ref{doppler}), the results are within expectations.

\begin{specialtable}[H]
\caption{\label{Variacao_doppler} Maximum and minimum variation of theoretical Doppler scattering.}
\tabcolsep=1.76cm
\begin{tabular}{ll}
\toprule 
\textbf{UAV Speed (km/h)} & \makecell{\textbf{Doppler Frequency} \\\textbf{(Minimum--Maximum)}} \\ 
\midrule 
1 & 0--0.86 Hz \\
3 & 0--2.6031 Hz \\
\bottomrule 
\end{tabular}
\end{specialtable}



To estimate the drone's speed, the level crossing rate was calculated in relation to its power level and then divided by the theoretical level crossing rate, therefore obtaining $f_{d}$ as a function of the signal strength. As the frequency used in communication is 915 MHz, its wavelength approximates 0.32~m. Using the equation $\hat{v} = \lambda\cdot\hat{f}_d/ \cos(\theta)$ (Subsection~\ref{sub_sec_doppler}), the speed is estimated. Figure~\ref{velocidade}a shows the estimated flight speed for the lake environment as a function of the angle $\theta$ = 0 $\degree$ (lowest speed). Figure~\ref{velocidade}b shows the speed estimation for the Caatinga and mixed environments. The same qualitative comments regarding the $f_d$ estimation are valid for the velocity estimation.


\clearpage
\end{paracol}
\nointerlineskip
\begin{figure}[H]
\widefigure
\begin{subfigure}{.5\textwidth}
  
  \includegraphics[width=1.0\linewidth]{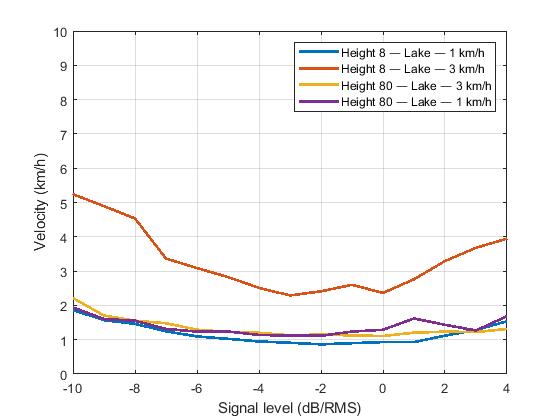}  
  \caption{}
  \label{Speed_lago}
\end{subfigure}~
\begin{subfigure}{.5\textwidth}
  
  \includegraphics[width=1.0\linewidth]{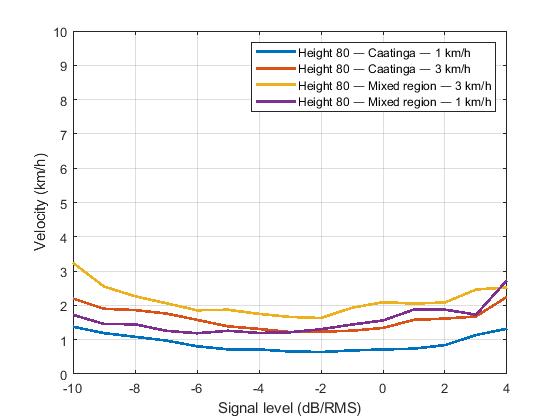}  
  \caption{}
  \label{Speed_outros}
\end{subfigure}

\caption{{Estimated speeds.} (\textbf{a}) Estimated speeds for Lake region. (\textbf{b}) Estimated speeds for the Caatinga and Mixed regions.}
\label{velocidade}
\end{figure}
\begin{paracol}{2}
\switchcolumn


\vspace{-6pt}
\section{Conclusions}\label{conclusion}

The contribution of this work was to characterize the air-ground channel used by a drone in three scenarios. Both small-scale and large-scale fading were estimated. Regarding large-scale fading, path loss and shadowing were characterized. At lower heights, the path loss presented a negative exponent, while a positive exponent was observed for higher heights. Shadowing parameters (mean and standard deviation) were also presented. We observed higher standard deviation for lower height flights. Thus, results show more severe long-scale fading for air-to-ground communication at lower altitudes, even over a lake surface.

We also present a small-scale fading study focused on comparing the theoretical and empirical CDFs of the collected samples. The results showed a higher variation in small-scale fading in the flight with the lower speed and lower height. We speculate that this result is related to the fact that the aircraft has less control of speed correction when it is affected by gusts of wind. The KS test shows that the Weibull distribution best describes the channel at a significance level of 95\%.

Finally, LCRs were used to estimate the Doppler frequency and drone speed. Even if the aircraft has its average speed strongly affected by wind, the expected results of the Doppler frequency were observed within the theoretical range. Flights over the lake at 3~km/h and a height of 8~m show better discrimination regarding speed and Doppler frequency estimations. This specific flight experiences a higher speed, because there was a barrier of vegetation on one of the lake's edges.

{It is worth noting that flights at low altitudes close to water presented a high path loss exponent, which can hinder the applications in this scenario. Such an effect should be better analyzed in order to comprehend the influence of the water surface on UAV communication channels at low altitude, as there is a lack of works in this subject area. Thus, we intend to investigate this in a future work.}

\vspace{6pt}
\authorcontributions{Conceptualization, D.L.L., V.A.d.S.J., and A.A.M.d.M.; Data curation, D.L.L.; Formal Analysis, D.L.L., V.A.d.S.J., and A.A.M.d.M.; Supervision, D.L.L., P.J.A., and A.A.M.d.M.; Visualization, D.L.L.; Writing---original draft, D.L.L.; Writing---reviewing \& editing, D.L.L., P.J.A., M.M.d.M.C., V.A.d.S.J., and A.A.M.d.M. All authors have read and agreed to the published version of the manuscript.}



\funding{This study was financed in part by the Coordena\c{c}\~{a}o de Aperfei\c{c}oamento de Pessoal de N\'{i}vel Superior---Brasil (CAPES)---Finance Code 001.}

\institutionalreview{{Not applicable. }
}

\informedconsent{{Not applicable.  }
}

\dataavailability{{Not applicable.  }
} 

\acknowledgments{{The proof of concept simulations provided by this paper was supported by the High Performance Computing Center at UFRN (NPAD/UFRN). }
}

\conflictsofinterest{The authors declare that they have no competing interests.}

\appendixtitles{no} 
\appendixstart
\appendix
\section{}
\unskip
\subsection{}
\label{ap1}

We use 4nec2 software to simulate the gain of the antenna lobe used by the XBee modules. The antenna radiation pattern is shown in Figure~\ref{antenna}. The antenna has its highest gain on the horizontal axis and its lowest gain on the vertical axis.

\begin{figure}[H]

\includegraphics[width=0.55\textwidth]{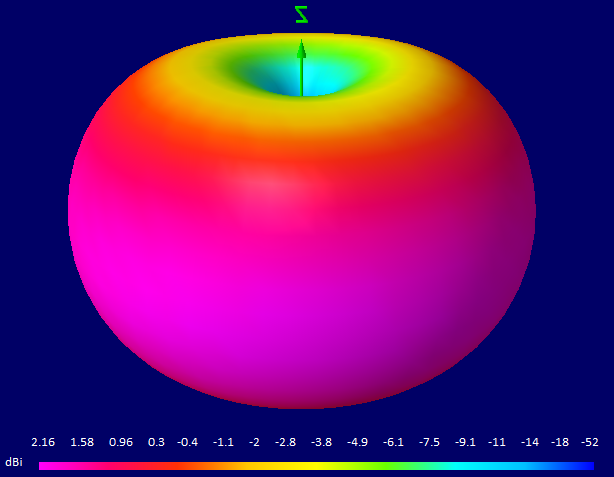}
\caption{\label{antenna} {XBee} 
 antenna lobe.}
\end{figure}

As illustrated in Figure~\ref{antenna_drones}, when the aircraft is at a higher height and close to the BS (overlapping antennas), the antennas provide low gains. Additionally, when they gain horizontal distance,
the main free Fresnel zone changes such that the antenna gain increases.

\begin{figure}[H]

\includegraphics[width=0.55\textwidth]{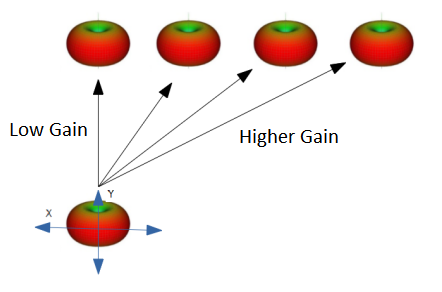}
\caption{\label{antenna_drones} Illustration of gain composition of the BS and drone antennas.}
\end{figure}



\end{paracol}
\reftitle{References}



\end{document}